\documentclass{jfm}

\usepackage{graphicx}
\usepackage{natbib}
\usepackage{caption}
\usepackage{subcaption}
\usepackage{cite}
\usepackage{amsmath}
\usepackage{color}

\ifCUPmtlplainloaded \else
  \checkfont{eurm10}
  \iffontfound
    \IfFileExists{upmath.sty}
      {\typeout{^^JFound AMS Euler Roman fonts on the system,
                   using the 'upmath' package.^^J}%
       \usepackage{upmath}}
      {\typeout{^^JFound AMS Euler Roman fonts on the system, but you
                   dont seem to have the}%
       \typeout{'upmath' package installed. JFM.cls can take advantage
                 of these fonts,^^Jif you use 'upmath' package.^^J}%
      }
  \else
  \fi
\fi

\ifCUPmtlplainloaded \else
  \checkfont{msam10}
  \iffontfound
    \IfFileExists{amssymb.sty}
      {\typeout{^^JFound AMS Symbol fonts on the system, using the
                'amssymb' package.^^J}%
       \usepackage{amssymb}%

      }{}
  \fi
\fi

\ifCUPmtlplainloaded \else
  \IfFileExists{amsbsy.sty}
    {\typeout{^^JFound the 'amsbsy' package on the system, using it.^^J}%
     \usepackage{amsbsy}}
    {\providecommand\boldsymbol[1]{\mbox{\boldmath $##1$}}}
\fi

\newcommand\Rey{\mbox{\textit{Re}}}  
\newcommand\As{A_\mathrm{s}}
\newcommand\at{a_\mathrm{trail}}
\newcommand\al{a_\mathrm{lead}}

\newsavebox{\astrutbox}
\sbox{\astrutbox}{\rule[-5pt]{0pt}{20pt}}

\DeclareMathOperator*{\argmin}{arg\,min}

\title[Viscous Conduit Solitary Wave Interactions]{Interactions of
  large amplitude solitary waves in viscous fluid conduits}

\author[N.~K. Lowman, M.~A. Hoefer, and G.~A. El]%
{N\ls I\ls C\ls H\ls O\ls L\ls A\ls S\ns K.\ns L\ls O\ls W\ls M\ls A\ls N$^1$%
	\thanks{Email address for correspondence: nklowman@ncsu.edu},\ns
M.\ns A.\ns H\ls O\ls E\ls F\ls E\ls R$^1$,\break\and 
G.\ns A.\ns E\ls L$^2$}

\affiliation{$^1$Department of Mathematics, North Carolina State University,
Raleigh, NC 27695, USA \\ [\affilskip] $^2$ Department of Mathematical Sciences, Loughborough University, Loughborough LE11 3TU, UK}

\date{?; revised ?; accepted ?. - To be entered by editorial office}
\begin{document}

\maketitle

\begin{abstract}
  The free interface separating an exterior, viscous fluid from an
  intrusive conduit of buoyant, less viscous fluid is known to support
  strongly nonlinear solitary waves due to a balance between
  viscosity-induced dispersion and buoyancy-induced nonlinearity.  The
  overtaking, pairwise interaction of weakly nonlinear solitary waves
  has been classified theoretically for the Korteweg-de Vries equation
  and experimentally in the context of shallow water waves, but a
  theoretical and experimental classification of strongly nonlinear
  solitary wave interactions is lacking.  The interactions of large
  amplitude solitary waves in viscous fluid conduits, a model physical
  system for the study of one-dimensional, truly dissipationless,
  dispersive nonlinear waves, are classified.  Using a combined
  numerical and experimental approach, three classes of nonlinear
  interaction behavior are identified: purely bimodal, purely
  unimodal, and a mixed type.  The magnitude of the dispersive
  radiation due to solitary wave interactions is quantified
  numerically and observed to be beyond the sensitivity of our
  experiments, suggesting that conduit solitary waves behave as
  ``physical solitons.''  Experimental data are shown to be in
  excellent agreement with numerical simulations of the reduced model.
  Experimental movies are available with the online version of the
  paper.
\end{abstract}


\section{Introduction}

Exponentially localized solitary waves arise in nature as a balance
between the steepening effects of nonlinearity and the spreading
effects of dispersion.  These fundamental, nonlinear coherent
structures exhibit an amplitude-dependent phase speed, often with
larger waves propagating faster than slower ones,
e.g. \citep{whitham_linear_1974}.  A distinguishing feature of
solitary waves in one-dimension is the nature of the resulting
interaction when a larger, trailing solitary wave overtakes a smaller,
leading wave, a scenario sometimes referred to as strong interaction
of solitary waves \citep{miles_obliquely_1977}.  A classical model of
weakly nonlinear solitary wave interactions is the Korteweg de-Vries
(KdV) equation.  Unlike in the case of linear waves, where
superposition applies, the nonlinear interaction is characterized by
the larger wave decreasing in height and experiencing a forward shift
in position, while the smaller solitary wave increases in amplitude
and experiences a negative position shift
\citep{zabusky_interaction_1965}.  The two solitary waves then emerge
from the interaction with their initial profiles and speeds restored.
The absence of interaction-induced radiation is associated with
mathematical solitons and complete integrability of the governing
equation, as is the case for KdV.  The exact two soliton
KdV solution was derived \citep{hirota_exact_1971} and the soliton
overtaking interation has been classified by amplitude ratio into
three distinct regimes according to the qualitative structure during
the interaction \citep{lax_integrals_1968}.  When the ratio of the
trailing and leading amplitudes is sufficiently small, a bimodal
structure persists through the interaction with the trailing wave
passing its mass forward to the leading wave.  When the ratio is
large, the interaction is unimodal, with the larger wave engulfing the
smaller one before emitting it.  For intermediate ratios, there is a
hybrid state, in which the interaction begins with the larger wave
absorbing the smaller one and forming an asymmetric, unimodal mass.
During the peak of the interaction, a distinctly bimodal wave appears
before the process undoes itself (see
Fig.~\ref{fig:interacts_numerics} for images of each interaction type
in the context of the model equation considered here).  This
classification scheme for KdV depends solely on the ratio of the
soliton amplitudes due to the existence of scaling and Galilean
symmetries.  It has been confirmed experimentally in the case of
weakly nonlinear, shallow water waves \citep{weidman_experiments_1978,
  craig_solitary_2006, li_amplification_2012}. Due to a capillary
instability and small dissipation, solitary water waves are limited to
nondimensional amplitudes less than $0.78$
\citep{tanaka_stability_1986,ablowitz_asymptotic_2010}, thus water
waves are a limited system in which to probe large amplitude,
conservative solitary wave interactions.  Moreover, experiments and
numerical simulations of the water wave equations accessed amplitudes
only up to $0.4$ \citep{craig_solitary_2006}.  Thus, to the authors'
knowledge, a systematic, quantitative classification of strongly
nonlinear solitary wave interaction behaviors in any physical system
is lacking experimentally and theoretically.

\begin{figure}
  \centering \includegraphics{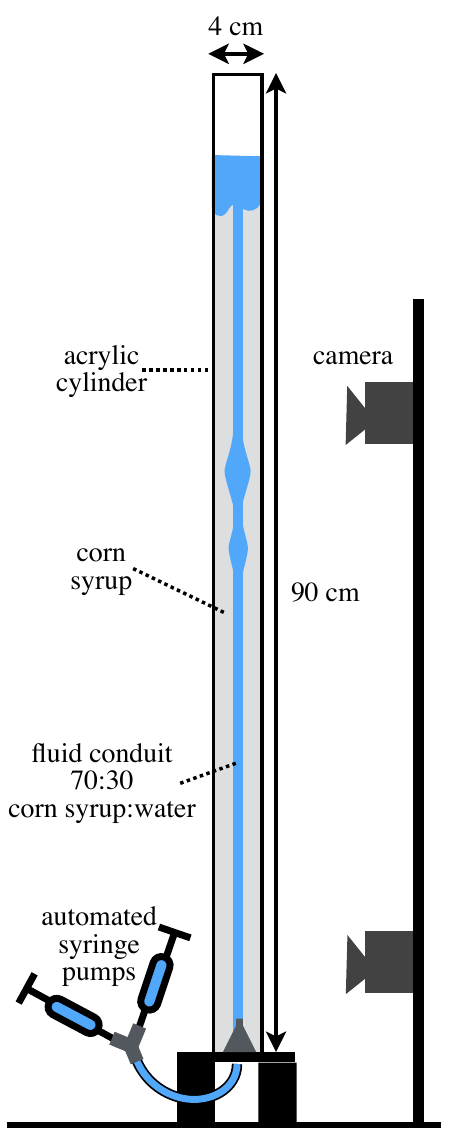}
  \caption{Diagram of the fluid conduit experimental system.}
  \label{fig:expt_schematic}
\end{figure}

In this work, we extend the classification of overtaking interactions
of solitary waves to a nonintegrable, strongly nonlinear,
dissipationless/dispersive wave equation, the so-called conduit
equation, cf.~\citep{lowman_dispersive_2013-1} and to experiments
involving solitary wave interactions with nondimensional amplitudes up
to $\sim 14$.  The conduit equation arises in the study of viscous
fluid conduits, a medium in which solitary waves have been studied
experimentally in isolation
\citep{scott_observations_1986,olson_solitary_1986} and
post-interaction \citep{helfrich_solitary_1990}, but not during the
interaction process.  The viscous fluid conduit setting is realized by
introducing a steady source of buoyant, viscous fluid to a quiescent
medium of heavier, more viscous fluid.  A stable, fluid-filled pipe is
formed.  Slow changes in the rate of injection induce interfacial
dynamics involving a maximal balance between buoyancy of the intrusive
fluid and the resistance to motion by the exterior fluid (see
Fig.~\ref{fig:expt_schematic}).  The scalar, nonlinear, dispersive
conduit equation capturing the interfacial dynamics has been derived
from the full set of coupled fluid equations
\citep{lowman_dispersive_2013-1}.  Unlike well-known models of small
amplitude, weakly nonlinear, interfacial fluid dynamics such as the
KdV \citep{korteweg_change_1895} and Benjamin-Ono
\citep{benjamin_internal_1967,ono_algebraic_1975} equations, the
conduit equation is derived under long wave assumptions only, valid
for large amplitudes \citep{lowman_dispersive_2013-1}, much like the
Green-Naghdi (or Serre, Su-Gardner) equations of large amplitude,
shallow water waves
\citep{serre_contribution_1953,su_kortewegvries_1969,green_derivation_1976}.
Moreover, large amplitude conduit solitary waves are asymptotically
stable \citep{simpson_asymptotic_2008}, exhibit good agreement with
experiments
\citep{scott_observations_1986,olson_solitary_1986,helfrich_solitary_1990},
and are robust, physical features of viscous fluid conduit interfacial
dynamics.

Using careful numerical simulations, we find that although the conduit
equation does not possess the KdV Galilean invariance, the qualitative
Lax classification scheme from KdV theory extends to the strongly
nonlinear regime for physically realizable solitary wave amplitudes.
The type of interaction depends on the absolute amplitudes of the two
waves, rather than solely on their ratio.  A scaling invariance of the
conduit equation renders a unit solitary wave background but cannot be
used to scale individual solitary wave amplitudes.  Our numerical
computations demonstrate small energy loss ($10^{-2}$ relative change
in the solitary wave two-norm) due to interaction, also numerically
observed in a closely related equation
\citep{barcilon_nonlinear_1986}.  This confirms the non-integrability
of the conduit equation, as shown by the Painlev\'{e} test
\citep{harris_painleve_2006}.  However, any dispersive radiation
following experimental solitary wave interaction was below the
resolution of our imaging system, a feature also observed in previous
experiments \citep{helfrich_solitary_1990}.  This suggests that while
not \textit{mathematical} solitons, conduit solitary waves are
\textit{physical} solitons.  We support these numerical observations
with quantitative interaction classification experiments, which are in
excellent agreement and represent the first observations of the mixed
and unimodal interaction types in viscous fluid conduits.

The importance of this work extends beyond the remarkable agreement
between theoretical and numerical predictions of conduit solitary wave
dynamics and our experimental observations.  In particular, the
overtaking interaction between two solitary waves can be seen as a
fundamental property of one-dimensional, dissipationless, dispersive
hydrodynamics.  As such, these observations further establish the
viscous fluid conduit setting as a practically accessible experimental
and theoretical platform for future investigations into solitary
waves, slowly modulated nonlinear wavetrains, and their interactions, for which
quantitative experiments in any physical system are essentially
lacking in the literature.  Moreover, the fact that we do not observe
qualitatively new behaviors in the interactions of solitary waves
beyond the weakly nonlinear regime is highly nontrivial due to the
lack of integrability and the increased dimensionality of the
parameter space.  This suggests there could be some robustness or
universality to the Lax categories for wave equations which
asymptotically reduce to KdV.  There is also renewed interest in the
nature of two soliton interactions in integrable and nearly integrable
systems in connection with the theory of soliton gas (or soliton
turbulence) \citep{el_kinetic_2005}.  Interactions falling into
different Lax categories have distinct effects on the statistical
characteristics of soliton turbulence
\citep{pelinovsky_two-soliton_2013}, and thus viscous fluid conduits
provide a promising setting for the experimental study of statistical
properties of incoherent soliton gases.

In the following section, we present the theoretical foundations for
the classification of conduit solitary waves and describe the
experimental set-up.  Section \ref{sec:classification} presents the
details of our findings, and the manuscript is concluded in
\S~\ref{sec:conclusions} with a discussion of future directions.

\section{Theoretical foundations} \label{sec:theoretical}

In this section, we give an overview of the theoretical foundations needed for classifying viscous fluid conduit solitary wave interactions.  We provide intuition regarding the physical effects that give rise to interfacial dynamics, develop relevant numerical tools, and describe the experimental set-up.


\subsection{Derivation of the conduit equation and solitary wave solutions}

A complete mathematical description of the viscous fluid conduit
setting requires the consideration of the full system of Navier-Stokes
equations for the intrusive and exterior fluids, with boundary
conditions along a moving, free interface.  To subvert this
difficulty, an approximate model governing the interfacial dynamics
has been derived from physical principles
\citep{olson_solitary_1986,scott_observations_1986} and via an
asymptotic, multiple scales procedure \citep{lowman_dispersive_2013-1}
using the ratio of the viscosities as the small parameter,
\begin{equation} 
	\epsilon = \mu^{(i)}/\mu^{(e)} \ll 1 \ ,
	\label{eq:epsilon}
\end{equation}
where $\mu^{(i)}$ indicates the (dynamic) viscosity of the intrusive
fluid and $\mu^{(e)}$ that of the exterior fluid.  Here we outline the
asymptotic derivation and identify the additional key nondimensional
quantities and assumptions required to ensure model validity for
comparison with experimental parameters.

The steady injection from below of a buoyant, viscous fluid into a basin of a much more viscous fluid evolves to form a vertically uniform, axisymmetric conduit, well-described by the governing equations of Poiseuille-type flow \citep{whitehead_dynamics_1975}.  In this unperturbed setting, the vertical velocity of the intrusive fluid is driven by a vertical pressure gradient due to buoyancy, and the conduit radius is set by the injection rate, assuming the velocity is $\mathcal{O}(\epsilon)$ at the interface.  In response to perturbations of the injection rate, radial velocities are excited along the interface, inducing a normal, viscous stress, which balances the pressure difference between the two fluids across the boundary.  This dominant balance is satisfied, provided the following assumptions hold: (1) the vertical variation along the conduit wall is small relative to the radial, i.e. the vertical length scale $L$ is large relative to the radius of the unperturbed conduit $R_0$,
\begin{equation}
	\label{eq:long_waves}
	L = \left(\frac{A_0}{8\pi\epsilon} \right)^{1/2} \ , \quad A_0 = \pi R_0^2 \ ,
\end{equation}
(2) the two fluids are miscible with negligible mass diffusion across the interface, and (3) the Reynolds number of the intrusive fluid, defined to be
\begin{equation}
	\Rey = \frac{\rho^{(i)}UL}{\mu^{(i)}} \ , \quad 
	U = \frac{gA_0 \Delta \rho}{8\pi \mu^{(i)} }\ ,
\end{equation}
for intrusive fluid density $\rho^{(i)}$ and density difference
$\Delta \rho = \rho^{(e)} - \rho^{(i)}$, is no larger than
$\mathcal{O}(1)$.  Under these provisions, the vertical pressure
gradient within the conduit has two contributions, one from buoyancy
and the other from viscous stress.  This leads to the
(nondimensional) volumetric flux $Q(z,t)$, with charactistic scale
$Q_0$, which can be written in terms of the nondimensional conduit cross-sectional area, $A(z,t)$ in the form
\begin{equation}
	\label{eq:flux}
	Q(z,t) = A^2(z,t) \left\{ 1 - \frac{\partial \ }{\partial z}
      \left[ \frac{1}{A(z,t)} \frac{\partial A(z,t) }{\partial t}
      \right] \right\} \ , \quad 
	Q_0 = A_0 U \ .
\end{equation}
Lastly, the flux is related to the evolution of the conduit area by
appealing to the continuity equation and imposing the kinematic
boundary condition along the interface, yielding $\partial_t
A(z,t)+ \partial_z Q(z,t) = 0$, or the conduit equation
\begin{equation}
	\label{eq:conduit} 
	\frac{\partial A(z,t)}{\partial t}  + \frac{\partial \ }{\partial
      z} \left( A^2(z,t) \left\{ 1 - \frac{\partial \ }{\partial z}
        \left[ \frac{1}{A(z,t)}  \ .\frac{\partial A(z,t) }{\partial
            t} \right] \right\} \right) = 0 \ . 
\end{equation}
The conduit equation can be approximated by the KdV equation in the
small amplitude, long wavelength regime
\citep{whitehead_korteweg-devries_1986}.  A key result for the present
study is that eq.~\eqref{eq:conduit} is valid for long
times, $t = \mathit{o}(\epsilon^{-1})$ and large amplitudes $A =
\mathit{o}(\epsilon^{-1})$, provided the aforementioned assumptions
are satisfied and $\epsilon$ is sufficiently small
\citep{lowman_dispersive_2013-1}.

Conduit solitary waves, first considered by \citep{scott_magma_1984},
are derived by introducing the ansatz $A(z,t) = A(\zeta)$, $\zeta =
z-ct$, where $c$ is the wave speed and $A$ decays exponentially to a
background constant, here taken to be unity without loss of
generality.  Inserting this form of the solution into
eq.~\eqref{eq:conduit} and integrating twice yields the ordinary
differential equation (ODE) defining the solitary wave profile
$A(\zeta)$, 
%
\begin{equation}
	\label{eq:solwave}
	\begin{split}
      & \alpha_0 \left( \frac{\mathrm{d} A}{\mathrm{d} \zeta}
      \right)^2 = \alpha_1 + \alpha_2 A + \alpha_3 A^2 + \alpha_4 A^2
      \ln{A} \ , \\
      & \quad \alpha_0 = \frac{1}{2} \left( 2\As^2\ln{\As} - \As^2+1
      \right) \ , \quad \alpha_1 = \As^2 \ln{\As} - \As^2+\As \ ,
      \quad \alpha_2 = -2 \alpha_0 \ , \\
      & \quad \alpha_3 = \As^2\ln{\As} - \As+1 \ , \quad \alpha_4 = -
      \left(\As^2-2\As+1\right) \ ,
	\end{split}
\end{equation}
where $\As$ is the total \emph{height} of the solitary wave,
i.e. background plus amplitude, and the amplitude-speed relation is
given by 
\begin{equation}
	\label{eq:speed}
	c(\As) = \frac{2\As^2\ln{\As} - \As^2+1}{\As^2-2\As + 1} \ .
\end{equation}
Note that the speed is monotone increasing in $\As$, so larger solitary waves always move faster.



\subsection{Numerical methods}

To study the overtaking interaction of conduit solitary waves, we
initialize $A$ in eq.~\eqref{eq:conduit} with two well-separated
solitary waves. The trailing wave has amplitude $a_\mathrm{trail}$ ($a
= \As-1$ is the amplitude above the background) and the lead wave has
amplitude $a_\mathrm{lead}$.  We take
$a_\mathrm{trail}>a_\mathrm{lead}$ so that
$c_\mathrm{trail}>c_\mathrm{lead}$.  The localized solitary waves are
separated initially so that their superposition on a uniform
background of unity exhibits small, $\mathcal{O}(10^{-7})$, difference
above background.  The ODE \eqref{eq:solwave} is integrated as in
\citet{lowman_dispersive_2013} with tolerance below
$\mathcal{O}(10^{-7})$.  The dynamical solver for \eqref{eq:conduit}
has been validated in \citet{lowman_dispersive_2013}.  The width of
the truncated spatial domain is chosen so that, at all times, the
solitary waves are $10^{-8}$ close to the background state at the end
points.  The grid spacing is chosen so that the individual solitary
waves are well-resolved, with values selected from the range $\Delta z
\in [0.05,0.5]$, with larger amplitude solitary waves requiring higher
resolution.  The time step is $\Delta t = \Delta z/2c_\mathrm{trail}$.


\subsection{Experimental set-up}

The experimental apparatus, depicted in Fig.~\ref{fig:expt_schematic},
used to study conduit solitary waves is an acrylic cylinder with
square sides 4~cm by 4~cm and a height of 90~cm, filled to a depth of
approximately 75~cm with a generic brand light corn syrup.  To ensure
miscibility, the intrusive fluid was taken to be a 70:30 mixture of
corn syrup and water, with food coloring used for imaging.  This
set-up closely follows previous experiments by
\citet{olson_solitary_1986,scott_observations_1986,helfrich_solitary_1990}.
Injection of the intrusive fluid through the base of the apparatus was
precisely controlled by use of an automated syringe pump, with the
base injection rate 0.1~mL/min to create a vertically uniform,
background conduit.  Solitary waves were formed by producing an
additional localized pulse in the rate of injection using a second
syringe pump, connected to the apparatus via a y-junction, hence
affording precision control on the size of the solitary waves
generated.  Viscosities of the two fluids were measured by a
rotational viscometer, with $2\%$ measurement uncertainty.  Densities
were measured using a scale and graduated cylinder with uncertainty
1\%.  Nondimensional, solitary wave amplitudes relative to background
were measured by counting pixels across the conduit from still frame
images captured with a digital SLR camera.  The dimensional radius of
the background conduit, held constant throughout the experiments, was
measured by comparing images of the background conduit with a grid of
known size attached to the back wall of the apparatus.  To compute the
correction due to the projection of the conduit fluid in the middle of the
apparatus onto the back wall, the grid was compared with a copy of the
same grid submerged within the filled apparatus before injection
commenced.  Errors due to imaging techniques and measurement were
estimated by measuring the diameter of the background conduit across a
range of images yielding a standard deviation of 2\%, on the order of
the viscosity measurements.  Interaction classification was achieved
by high definition video recording of the interaction using a second
camera.  Still frames of the interactions were then extracted from the
video, and downsampled using bicubic interpolation in the vertical
coordinate by a factor of $0.1 \approx \epsilon^{1/2}$ in order to
enforce an aspect ratio of 1.  Recall the long wavelength scaling in
\eqref{eq:long_waves} sets an aspect ratio of the vertical to radial
lengths of order $\epsilon^{-1/2}$.  This scaling significantly
improves the fidelity with which we can classify the solitary wave
interaction types.  In cases where it was difficult to determine the
classification, edge detection algorithms were also used.  Measured
and derived fluid properties are provided in Table \ref{tab:dimless}.

A major difficulty previously encountered during experiments with this
system was creating and maintaining a straight, vertical conduit.  We
find the following protocol to be effective.  The injection line is
prepared so that a small amount of air is left in the line just ahead
of the intrusive fluid.  The remaining intrusive fluid has no air
bubbles.  A well-mixed volume of corn syrup is poured down the side of
the cylinder to fill, minimizing the entrainment of air.  The
apparatus is allowed to equilibrate overnight.  The experiment is
initiated with steady injection at a rate of 0.5 mL/min.  First,
controlled air bubbles are produced so that the initial penetration of
the intrusive fluid follows behind the air bubbles.  This latter
protocol is similar to the procedure described in
\citet{helfrich_solitary_1990}.  We find the background conduit to be
straight to within $0.2^\circ$ across 60 cm.  It merits mention that
the conduit equation \eqref{eq:conduit} has been shown to be valid for
conduits canted by $\mathcal{O}(6^\circ \approx
\epsilon^{1/2}180^\circ/\pi)$ or less
\citep{lowman_dispersive_2013-1}, which was not violated here due to
our controlled initiation procedure.

\begin{table}
  \begin{center}
\def~{\hphantom{0}}
  \begin{tabular}{c|c|c|c|c|c|c|c} 
      $\rho^{(i)}$   & $\rho^{(e)}$  &   $\mu^{(i)}$  & $\mu^{(e)}$  & $A_0$   &   $U$ & $\Rey$ & $\epsilon$ \\[3pt] 
      1.23  g/mL & 1.37 g/mL & 0.789 P & 83.6 P &
        0.017 cm$^2$ & 0.118 cm/s & 0.049 & $9.4 \times 10^{-3}$ \\
  \end{tabular}
  \caption{Key experimental parameters.}
  \label{tab:dimless}
  \end{center}
\end{table}


\section{Overtaking interactions between strongly nonlinear solitary waves} \label{sec:classification}

\begin{figure}
  \centering \includegraphics{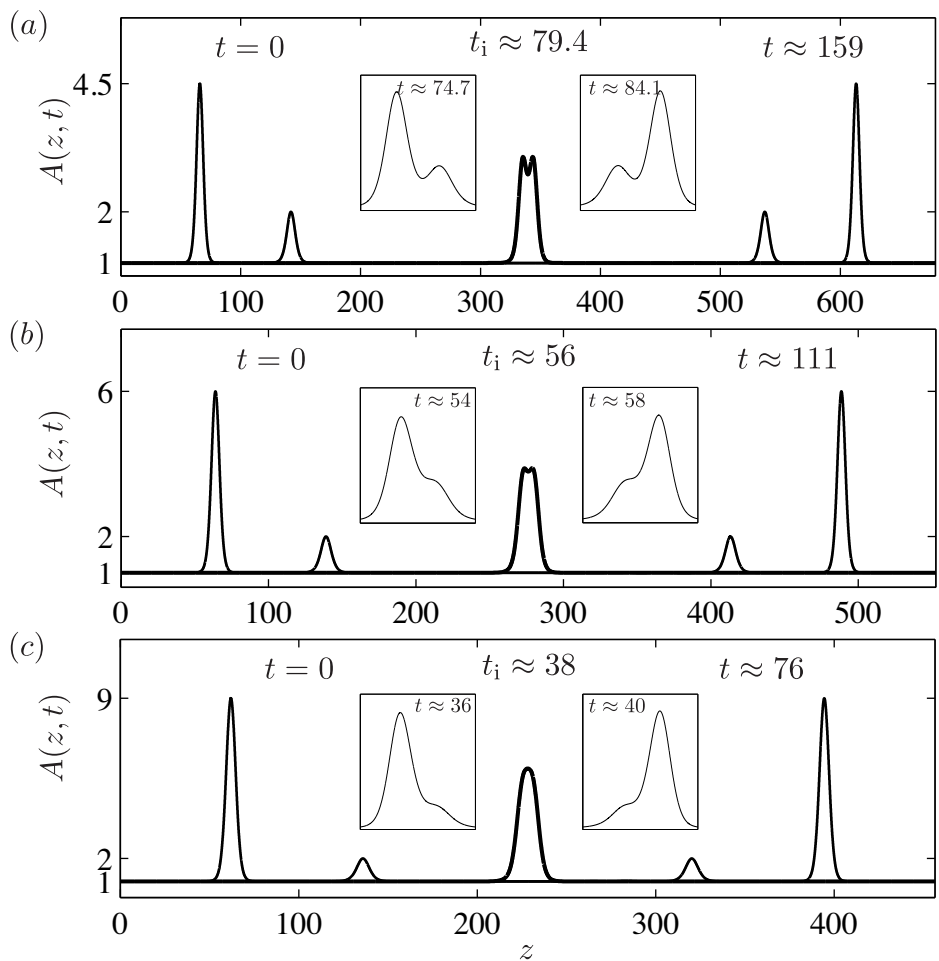}
  \caption{Example numerical solutions of the conduit equation
    eq.~\eqref{eq:conduit} exhibiting the three types of overtaking
    interactions.  The initial and final states, as well as the
    solution at the time of peak interaction $t_\mathrm{i}$, are
    plotted on the spatial axis while the left and right insets
    correspond to the solution just before and just after the peak
    interaction.  The trailing amplitude is varied, while the leading
    amplitude is fixed, $a_\mathrm{lead}=1$.  $(a)$ Bimodal
    interaction, $a_\mathrm{trail} = 3.5$. $(b)$ Mixed interaction,
    $a_\mathrm{trail} = 5$.  $(c)$ Unimodal interaction,
    $a_\mathrm{trail} = 8$. }
  \label{fig:interacts_numerics}
\end{figure}

Using the theoretical, experimental, and numerical techniques developed in the previous section, we now describe the classification of strongly nonlinear solitary wave interactions in the viscous fluid conduit setting.  Long time, high resolution numerical simulations in Fig.~\ref{fig:interacts_numerics} exhibit the three interaction categories, which are also found experimentally and displayed in a photo montage in Fig.~\ref{fig:expt_images}.  It is further shown that the dispersive tail generated by solitary wave interactions is beyond the sensitivity of our experiments.  


\subsection{Classification of interactions:  KdV}

In the case of the KdV equation, i.e. the weakly nonlinear, long
wavelength regime, properties due to integrability have been used to
classify the overtaking interaction analytically into three distinct
categories, based solely on amplitude ratio $\at/\al$
\citep{lax_integrals_1968}:
\begin{equation}
  \label{eq:1}
  \begin{split}
    1 < \frac{\at}{\al} < \frac{3+\sqrt{5}}{2} \approx 2.62 \ &: \quad
    \text{bimodal} \ ,\\
    \frac{3+\sqrt{5}}{2} < \frac{\at}{\al} < 3 \ &: \quad \text{mixed} \ , \\
    \frac{\at}{\al} > 3 \ &: \quad \text{unimodal} \ ,
  \end{split}
\end{equation}
where $\at$ and $\al$ are the trailing and leading soliton amplitudes,
respectively for $t \to -\infty$.  A bimodal interaction denotes the
case where the wave complex maintains a bimodal structure throughout
the interaction.  This type of exchange interaction corresponds to a
transfer of mass from the larger, trailing solitary wave to the
smaller, lead solitary wave. In contrast, unimodal interaction
involves the complete fusion of the lead wave by the trailing wave,
followed by fission into two waves.  The intermediate, mixed-type
interaction, which has a limited range of amplitude ratios in the
weakly nonlinear case, possesses both qualities, a unimodal structure
just before and just after interaction but a distinctly bimodal one at
$t = t_\mathrm{i}$.

\subsection{Classification of interactions:  conduit equation}
\label{sec:class-inter-cond}

The numerical classification of strongly interacting solitary waves is
achieved by dynamical evolution of the conduit equation
\eqref{eq:conduit} given initial data consisting of a large, trailing
solitary wave of amplitude $a_\mathrm{trail}$, well separated from a
smaller, leading solitary wave of amplitude $a_\mathrm{lead}$.  The
geometry of the wave structure near the time of interaction,
$t_\mathrm{i}$, defined to be
\begin{equation}
  \label{eq:2}
  t_\mathrm{i}= \argmin_{t} \left\{ \max_{z} \left[ A(z,t) \right]
  \right\} \ ,
\end{equation}
is examined.  The choice of $t_\mathrm{i}$ is due to the nature of the
interaction, in which the larger solitary wave decreases in amplitude
until $t = t_\mathrm{i}$ before asymptotically approaching
$a_\mathrm{trail}$ long after the interaction.  This behavior is
characteristic of KdV soliton interactions as well.

We allow the two solitary wave initial profiles to evolve long past
the time of interaction ($t_\mathrm{final} \approx 2t_\mathrm{i}$).
Once the numerical solution is obtained, the location and height of
the two solitary wave maxima are obtained for each time step by
interpolating the solution onto a finer grid and examining its
derivative to find the local extrema.  If only one maximum is found,
the structure at that time is considered unimodal.  We classify
interactions as bimodal if two maxima are present throughout and as
unimodal if the interaction possesses only one peak at $t =
t_\mathrm{i}$.  Note that the distinguishing feature of the mixed
interaction is the presence of a single maximum just before and after
the peak interaction time, but the reemergence of two distinct maxima
at $t = t_\mathrm{i}$.  Example numerical simulations of
eq.~\eqref{eq:conduit} of each type of interaction for a fixed $\al =
1$ and varying $\at$ are presented in
Fig.~\ref{fig:interacts_numerics}.

\begin{figure}
	\centering \includegraphics{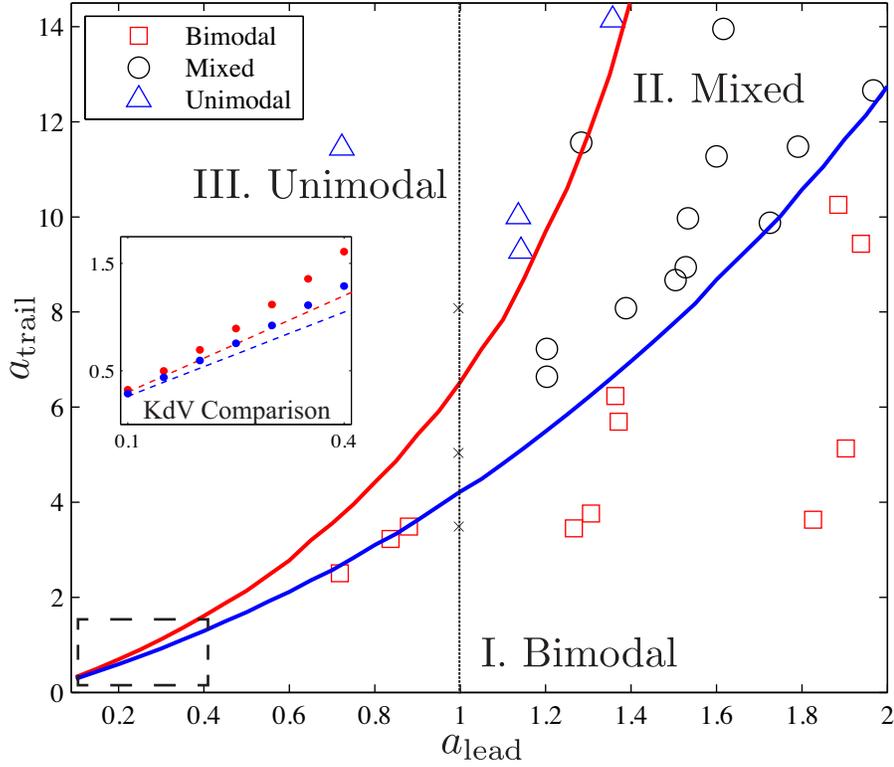}
	\caption{Phase diagram of the numerical and experimental
      classification of the overtaking interaction between two conduit
      solitary waves as a function of the leading and trailing wave
      amplitudes.  The blue (darker) curve indicates the transition
      from bimodal interaction to the intermediate/mixed type.  The
      red (lighter) curve marks the transition from the mixed type to
      unimodal.  The inset represents the boxed portion in the weakly
      nonlinear regime showing convergence to the KdV predictions
      (dashed lines).  Crosses along $a_\mathrm{lead} = 1$ correspond
      to simulations in Fig.~\ref{fig:interacts_numerics}.  The
      geometric shapes correspond to experimental classification.}
	\label{fig:bifurcation}
\end{figure}

The bifurcation diagram in Fig.~\ref{fig:bifurcation} presents the key
results of our classification analysis.  For a range of leading and
trailing amplitude solitary waves, the critical amplitudes marking
phase transition are plotted.  This was determined by fixing $\al$ and
monotonically varying $\at$ in increments of $\al/20$ until the
interaction type had transitioned from one type to another for three
consecutively larger values of $\at$.  The critical value then was
taken to be the value of $\at$ midway between the amplitudes
corresponding to the last interaction of one type and the first
interaction of the new type.  The x marks on the dashed vertical line
along $\al = 1$ mark the location of the simulations presented in
Fig.~\ref{fig:interacts_numerics}.  We find that due to a continuous
transition, the precise determination of type I-III requires high
resolution simulations.  

As pointed out earlier, the behavior of the conduit equation
\eqref{eq:conduit} is asymptotically equivalent to KdV in the small
amplitude regime, which is captured in the zoomed inset of the phase
diagram as the conduit transitions limit on the KdV transitions for
sufficiently small $\at \lessapprox 0.5$.  However, in this
nonintegrable, strongly nonlinear equation where $2 < \at < 15$, the
type of interaction depends not on the amplitude ratio, but on the
values of both amplitudes.  This is due to the existence of three
distinct conduit amplitudes, the background and those of the trailing
and leading solitary waves.  Only one amplitude can be scaled to unity
using symmetry of the equation, leaving two other free parameters
(cf. \citet{lowman_dispersive_2013}).

The complete, mathematical classification of KdV soliton interactions
was enabled by an explicit representation of the solution.  Here, we
do not have this luxury.  Like in the integrable setting, though, the
structure of the interaction for every amplitude tested in our
simulations (which covers most of the physically relevant range)
always falls into one of the three types.  Moreover, the mixed
geometry is expected for a much wider range of amplitudes than in
\eqref{eq:1} as the two initial waves grow larger.

\subsection{Radiation emitted due to interaction}

It is also of physical interest to consider the magnitude of the
dispersive tail resulting from interactions of conduit solitary waves,
which are not exact solitons.  Overtaking interactions of solitary
waves in nonintegrable equations have been shown via numerical
simulations to produce a small tail of linear dispersive waves
following their interaction,
e.g. \citep{bona_solitarywave_1980,mirie_collisions_1982,barcilon_nonlinear_1986},
a feature which if sufficiently large, could be examined
experimentally.  To address this issue, we have run simulations of
solitary wave interactions for a fixed $\al = 1$ and $\at$ varying
between 2 and 8, so that it spans all three interaction types and also
corresponds to the experiments in the following section.  The
radiation was quantified in two ways using long time numerical
evolution, $t_\mathrm{final} \approx 3 t_\mathrm{i}$.  The first is
the change in the amplitudes of the solitary waves post-interaction
and the second is the change in the profiles.  Here we find the
maximum change in amplitudes for both waves is consistently
$\mathcal{O}(10^{-3})$.  The change in the individual solitary wave
profiles is determined by centering a window about each individual
wave, for both the initial and final times, and then determining the
residual between the two profiles, here defined by the relative
two-norm difference, i.e. two-norm of the residual divided by the
initial two-norm.  This metric reveals that the change in profiles
from before to long after the interaction is not larger than
$\mathcal{O}(10^{-2})$ across the simulations examined.  These
findings are consistent not only with numerical simulations of a
closely related equation \citep{barcilon_nonlinear_1986}, but also
with experimental findings from conduit solitary wave interactions
\citep{helfrich_solitary_1990}.  Moreover, the amplitude differences
and residuals are beyond the sensitivity of our experimental
capabilities, which suggests that these conduit solitary waves are
approximately solitons, at least in a physical sense, hence we term
them ``physical solitons.''

%


\subsection{Experimental observation of the three types of interaction}

\begin{figure}
	\centering \includegraphics[width=\columnwidth]{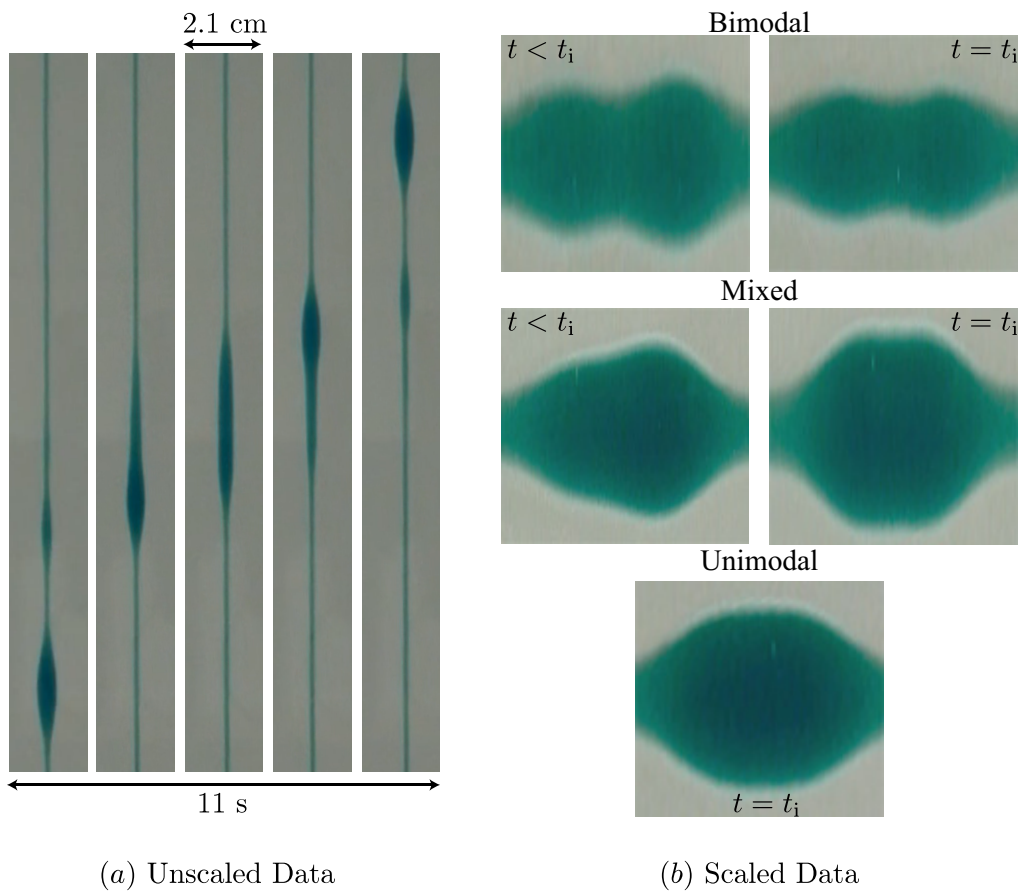}
	\caption{Data from interaction experiments. $(a)$ Unscaled data
      showing the solitary wave profiles from a typical interaction.
      $(b)$ Examples of the three interaction types for scaled data
      used in classifications.  From top to bottom,
      $(a_\mathrm{lead},a_\mathrm{trail})$ are $(1.27,3.45)$,
      $(1.53,9.97)$, and $(1.14,10.01)$, respectively.  See the online
      version of the paper for movies of these experiments.}
	\label{fig:expt_images}
\end{figure}

In Fig. \ref{fig:bifurcation}, we plot the results of twenty-seven
solitary wave interaction classification experiments.  The three
distinct types predicted by numerical simulations of the conduit
equation \eqref{eq:conduit} are readily observable in the full
physical system, and their dependence on $a_\mathrm{lead}$ and
$a_\mathrm{trail}$ is in excellent agreement with the phase diagram.
Example images of an unscaled interaction experiment and then scaled
data used for classification are given in Fig.~\ref{fig:expt_images}.
While it is sometimes difficult to distinguish between the regimes in
the unscaled data, scaling the data recovers the aspect ratio of the
nondimensional coordinate system from the numerical simulations and
allows for proper determination.  Typical examples of the three
interaction types are shown.  Movies in both unscaled and scaled
formats are available with the online version of the paper.

Regarding the agreement with numerical simulations, up to the 2\%
error in measuring conduit diameters, the data all lie in the
appropriate regions of the phase diagram.  While this agreement is
compelling on its own, it is also possible to compare with the
approximate model breakdown criteria identified in
\citet{lowman_dispersive_2013-1}.  The use of miscible fluids meets
the negligible surface tension criterion, and the contributions due to
the outer wall are small because the nondimensional radius of the
outer wall satisfies $R_\mathrm{wall} \approx 76 >> \epsilon^{-1/2} =
10.3$.  The Reynolds number criterion $Re = 0.049 \ll \epsilon^{-1} =
106$ for neglecting inertial effects is also satisfied.  Lastly, the
breakdown of the multiple scales assumption occurs when the condition, $A \ll 1/8\epsilon \approx 13.3$, is violated for large amplitude solitary waves \citep{lowman_dispersive_2013-1}.  We note that two of the trials lie beyond this point, though they still fall in the appropriate classification region.  There exists an independent condition for model validity based on the introduction of inertial effects for sufficiently large amplitude solitary waves, measured by an effective solitary wave Reynolds number \citep{helfrich_solitary_1990}, but we find the multiple scales condition to be more restrictive in our case.  Our results suggest remarkably robust concurrence between the reduced,
approximate interfacial equation and the full two-fluid system.


\section{Summary and conclusions} \label{sec:conclusions}

The qualitative characterization of large amplitude, pairwise solitary
wave interactions in viscous fluid conduits has been shown to permit
geometric classification according to the three Lax categories for
KdV.  Unlike the weakly nonlinear regime, however, the expected
interaction type depends on the wave amplitudes, rather than only
their ratio, and the mixed unimodal/bimodal interaction type is a more
robust, readily observable feature than for surface water waves.

The long-time, large amplitude validity of the conduit equation
\eqref{eq:conduit} and its analytical tractability make this two
viscous fluid setting an ideal one for the study of nonlinear
dispersive waves.  That nonlinear dispersive waves occur at all in a
fully viscous setting is a nontrivial observation, but that the
reduced equation captures the geometry of interacting solitary waves
suggests the interfacial dynamics of viscous fluid conduits are, as
predicted, approximately one-dimensional and dissipationless at the
time scales under consideration.  Moreover, the absence of dispersive
radiation in the experiments implies that, while the conduit equation
is not completely integrable, its solitary waves practically interact
elastically.  These results encourage future experimental studies on
nonlinear coherent structures, such as rarefaction waves, slowly
modulated wavetrains (dispersive shock waves) and their
interactions.


\section*{Acknowledgements}
We thank Karen Daniels, Hien Tran, and the NCSU Department of Mathematics for laboratory support.  This work was funded by NSF Grant Nos. DGE-1252376, DMS-1008793, and CAREER DMS-1255422. 

\bibliographystyle{jfm}

\end{document}